\documentclass[runningheads]{llncs}
\usepackage{enumitem}
\usepackage{graphicx}
\usepackage{algorithm}
\usepackage{algorithmicx}
\usepackage[noend]{algpseudocode}
\usepackage{wrapfig, epsfig}
\usepackage{makecell}
\usepackage{amsmath}

\newlength{\bibitemsep}
\newlength{\bibparskip}
\let\oldthebibliography\thebibliography
\renewcommand\thebibliography[1]{%
  \oldthebibliography{#1}%
  \setlength{\parskip}{\bibitemsep}%
  \setlength{\itemsep}{\bibparskip}%
}
\usepackage[nodisplayskipstretch]{setspace}
\begin{document}
\title{Fairness-aware Crowdsourcing of IoT \\ Energy Services}

\author{
Abdallah Lakhdari
\and
Athman Bouguettaya
}
\authorrunning{A. Lakhdari et al.}
%
\institute{
School of Computer Science, University of Sydney, Australia\\
\email{\{abdallah.lakhdari,athman.bouguettaya\}@sydney.edu.au}}

\maketitle              

\begin{abstract}
We propose a Novel Fairness-Aware framework for Crowdsourcing Energy Services (FACES) to efficiently provision crowdsourced IoT energy services. Typically, efficient resource provisioning might incur an unfair resource sharing for some requests. FACES, however, maximizes the utilization of the available energy services by maximizing fairness across all requests. We conduct a set of preliminary experiments to assess the effectiveness of the proposed framework against traditional fairness-aware resource allocation algorithms. Results demonstrate that the IoT energy utilization of FACES is better than FCFS and similar to Max-min fair scheduling. Experiments also show that better fairness is achieved among the provisioned requests using FACES compared to FCFS and Max-min fair scheduling.
\end{abstract}
\vspace{-20pt}
\section{Introduction}
The proliferation of the Internet of things (IoT)  may give rise to a self-sustained crowdsourced IoT ecosystem \cite{atzori2010internet}. The augmented capabilities of IoT devices such as sensing and computing resources may be leveraged for peer-to-peer sharing. People can exchange a wide range of IoT services such as computing offloading, hotspot proxies, {\em energy sharing}, etc. These crowdsourced IoT services present a convenient, cost-effective, and sometimes the only possible solution for a resource-constrained device \cite{ahabak2015femto}. For instance, a passenger's smartphone with low battery power may elect to receive energy from nearby wearables {\em using Wifi}\cite{raptis2019online}. The focus of this paper is on crowdsourcing IoT energy services. 

The concept of {\em wireless energy crowdsharing} has been recently introduced to provide IoT users with power access, anywhere anytime, through crowdsourcing \cite{bulut2018crowdcharging}\cite{raptis2019online}\cite{Previouswork11}. We leverage the service paradigm to unlock the full potential of IoT energy crowdsourcing. We define {\em an IoT Energy Service} as the abstraction of energy wireless delivery from an IoT device (i.e., {\em provider}) to another device (i.e., {\em consumer}) \cite{lakhdari2020composing}. Crowdsourcing IoT energy services has the potential of creating a \textit{green} service exchange environment by \textit{recycling} the unused IoT energy or relying on \textit{renewable} energy sources. For example, an IoT device may share its spare energy with another IoT device in its vicinity. Another example, a smart shoe may harvest energy from the physical activity of its wearer~\cite{choi2017wearable}\cite{gorlatova2014movers}. Additionally, wireless charging allows energy crowdsharing to be a \textit{convenient} alternative as the devices do not need to be tethered to a power point, nor use power banks.  Crowdsourcing energy services can be deployed through already existing wireless power transfer technologies such as \textit{Energous}\footnote{https://www.energous.com/}  that can deliver up to 3 Watts power within a 5-meter distance to multiple receivers. 
\begin{figure}[!t]
\centering
\includegraphics[width=0.5\textwidth]{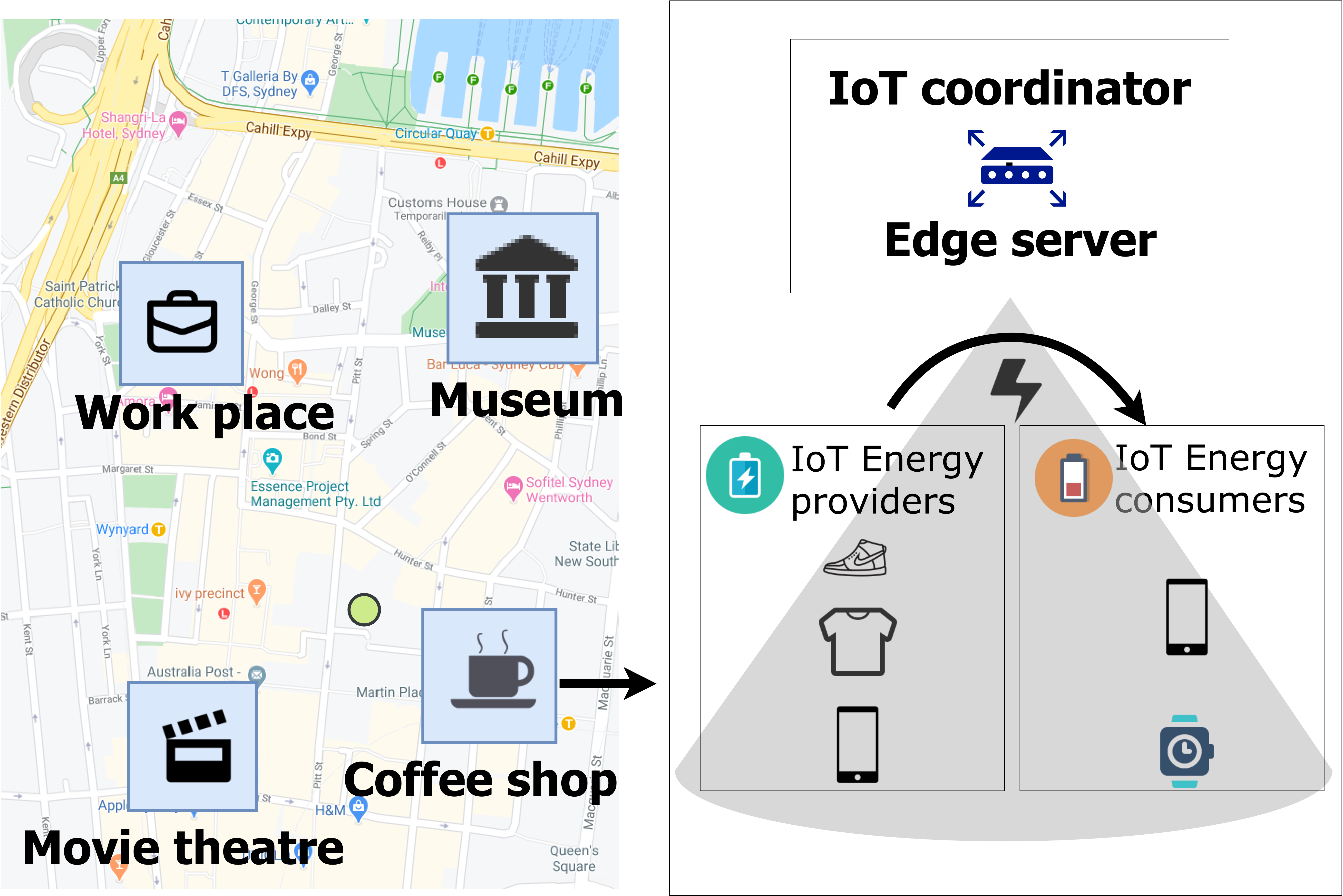}
\setlength{\abovecaptionskip}{0pt}
\setlength{\belowcaptionskip}{-18pt}
\caption{\small Crowdsourcing IoT energy services}
\label{fig:MicrocellSharing}
\end{figure}

The crowdsourced IoT energy ecosystem is a \textit{dynamic} environment that consists of providers and consumers congregating and moving across \textit{microcells} boundaries. A microcell is any confined area in a smart city where people may gather (e.g., coffee shops, restaurants, museums, libraries), see Fig. \ref{fig:MicrocellSharing}. The deployment of the energy crowdsharing ecosystem depends on the willingness of the IoT device owners to participate. Indeed, providers may share their energy {\em altruistically} to contribute to a {\em green} IoT environment. They may also be motivated by  {\em egotistic} purposes where participants are encouraged to share energy through a set of {\em incentives} \cite{wu2017revenue}. We assume that the IoT coordinator provides incentives to encourage energy sharing in the form of credits. These would be used to receive more energy when the providers act as consumers in the future \cite{wu2017revenue}\cite{abusafia2020reliability}\cite{abusafia2020incentive}. The IoT coordinator is assumed to be deployed one hop away from the energy providers and consumers (e.g., router at the edge) to minimize the communication overhead and latency while advertising energy services and requests. The participation of IoT users in the energy crowdsharing ecosystem depends on the security and trust of the deployed ecosystem. Novel security modules and new privacy-preserving trust models have been developed for crowdsourced IoT environments. These aspects are outside the scope of this paper. Our primary focus in this work is on fairness-aware crowdsourcing of IoT energy services.

We propose a fairness-aware service provisioning framework to cater for {\em multiple} energy requests in a crowdsourced IoT market. \textit{The under-provision of energy requests may demotivate consumers to participate in the crowdsourced IoT energy market}. In this paper, we focus on the notion of {\em fairness} in provisioning IoT energy services to satisfy the maximum number of energy requests. Sometimes, in a crowdsourced IoT environment, the available energy services in a microcell may not satisfy all existing requests. It is challenging to satisfy consumers by fulfilling only parts of their energy requirements. An efficient scheduler, in traditional resource allocation algorithms, aims at maximizing the throughput (i.e., the amount of resource utilization in a unit of time) \cite{graham1979optimization}. In some embedded systems, the scheduler must also ensure meeting deadlines of multiple requests \cite{cheng2004concise}. Typically, maximizing the throughput and meeting requests' deadlines might incur an unfair resource sharing for some requests (e.g., starvation of long requests in a short job first scheduler). In a crowdsourced IoT energy market, however, we claim that {\em if more energy requests are satisfied with respect to their time intervals (i.e., fairness), more energy would be consumed}. We transform the fairness-aware service provisioning problem into an optimization problem, i.e., maximizing the utilization of the available energy services by maximizing fairness across all requests. The contributions of this paper are:
\begin{itemize}[itemindent=0pt]
    \item[$\bullet$] A formulation of the IoT energy services provisioning problem as a time-constrained optimization problem.
    \item[$\bullet$] A fairness model to accommodate multiple IoT energy requests in the crowdsourced IoT environment.
    \item[$\bullet$] A spatio-temporal framework for fairness-aware crowdsourcing of IoT energy services (FACES).
    \item[$\bullet$] An experimental analysis with two implementations of the proposed fairness-aware energy crowdsourcing framework.
\end{itemize}

\begin{figure}[!t]
\centering
\setlength{\abovecaptionskip}{-1pt}
\setlength{\belowcaptionskip}{-18pt}
 \includegraphics[width=0.85\textwidth]{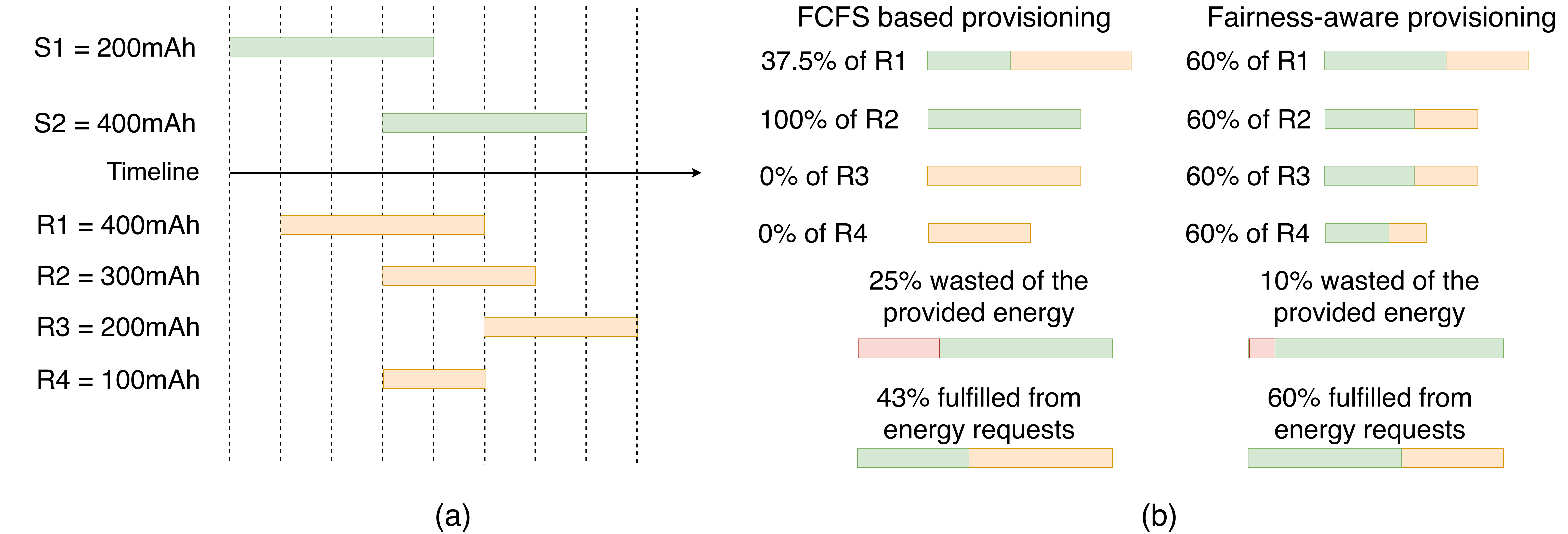}
\caption{\small (a) Time constraints of energy services and requests (b) Energy provisioning}
    \label{fig:MotivScen}
\end{figure}
\vspace{-20pt}
\section{Motivating scenario}
\vspace{-5pt}

We will use the following scenario: Six IoT users staying at a coffee shop. Two users are willing to provide their energy services $S_1$ and $S_2$. The other four IoT users are requesting energy from their neighboring IoT devices $R_1$, $R_2$, $R_3$, and $R_4$. The advertisement of services and requests includes various information, e.g., the start time, the end time, the location of the IoT device, and the provided or requested energy amount. Fig. \ref{fig:MotivScen} (a) illustrates the available services and requests by their timelines. It also shows the amounts of provided and requested IoT energy. It is challenging to allocate energy and fulfill multiple requests' requirements when there are limited energy services. For example, in Fig. \ref{fig:MotivScen} the available energy represents only $60\%$ of the total requested energy. 


Fig. \ref{fig:MotivScen} (b) presents the outcome of different allocation plans for the available energy to $R_1$, $R_2$, $R_3$, and $R_4$. Our goal is to fairly and efficiently allocate energy services to all requests. The first allocation plan follows the FCFS (i.e., First Come First Served) scheduling strategy. Each request is provided with energy according to its arrival (i.e., start time) and the available energy at its time interval. For example, in Fig. \ref{fig:MotivScen} (a)
$R_1$ receives energy from $S_1$. However, when $R_2$ arrives, it receives energy only from $S_2$. Even $S_1$ is also within its time interval, $S_1$ has been already reserved by $R_1$. Scheduling strategies such as FCFS and priority-based schedulers may not be a good fit to provision crowdsourced IoT energy. These strategies fulfill the requirements of each energy request independently and sequentially based on their arrival time. Services may fulfill the requirements of an energy request without being fully utilized, which affects the energy allocation efficiency. For example, all the requirements of $R_2$ have been fulfilled. On the other hand, $R_3$ and $R_4$ did not receive any energy. The FCFS-based scheduling could fulfill only $43\%$ of the total amount of the requested energy and wasted $25\%$ of the total available IoT energy.

The coordinator aggregates the provided energy by all services based on their time intervals. The aggregated energy is then shared among all  requests according to their time constraints. The second energy allocation plan in Fig. \ref{fig:MotivScen} (b) is a good illustration of the effect of a fairness-aware provisioning plan. Allocating $60\%$ of the requirement to each energy request maximizes the consumption of the available energy to $90\%$ (i.e., $10\%$ wastage). The limited provided energy and the time constraints of requests represent critical challenges for efficient and fairness-aware provisioning of IoT energy services. We reformulate our service provisioning problem as {\em a multi-objective time-constrained optimization problem, i.e., (i) maximizing the allocated energy provided by the available services, and (ii) maximizing the fairness for each request with respect to its time constraints}.


\vspace{-5pt}
\section{Preliminaries}\label{sysmdl1}

We first adopt the definitions of IoT energy services and requests in \cite{Previouswork11}. We then introduce the concept of fairness among energy requests based on their allocated energy. This work considers a provisioning framework for stationary services and requests to focus only on the temporal constraints in allocating energy to multiple requests in a microcell within a predefined time interval. The goal is to ensure fairness over a predefined time while maximizing green energy provision. In the future, we will extend the framework to fit into a  dynamic crowdsourced market by dealing with moving services and requests.



\paragraph*{Definition 1}
\textit{An energy service $ CES $} is a tuple $< Eid, Eownerid, F , Q >$ where:
\begin{itemize}[ noitemsep,nosep,leftmargin=10pt,labelsep=1pt,itemindent=0pt, labelwidth=*]
\item[--] $Eid$ is a unique service ID, 
\item[--] $Eownerid$ is a unique ID for the owner of the IoT device, 
\item[--] $F$ is the set of $CES$ functionalities offered by an IoT device $D$. 
\item[--] $Q$ is a tuple of $<q_1, q_2, ..., q_n>$ where each $q_i$ denotes a QoS property.
\end{itemize}
\paragraph*{Definition 2}
\textit{Crowdsourced IoT energy  Quality of Service (QoS) Attributes} allow users to distinguish among crowdsourced IoT energy services. QoS parameters are defined as a tuple $< l, St,Et, DEC,I, Tsr, Rel_i>$ \cite{lakhdari2020composing} where: 
\begin{itemize}[ noitemsep,nosep,leftmargin=10pt,labelsep=1pt,itemindent=0pt, labelwidth=*]
    \item[--] $l$ is the location of the provider. 
    \item[--] $St$ represents the start time of a crowdsourced IoT energy service.
    \item[--] $Et$ represents the end time of a crowdsourced IoT energy service respectively. 
    \item[--] $DEC$ is the deliverable energy capacity. 
    \item[--] $I$ is the intensity of the wirelessly transferred current. 
    \item[--] $Tsr$ represents the transmission success rate. 
    \item[--] $Rel_i$ represents the reliability QoS
\end{itemize}

The spatio-temporal features of the IoT energy services (i.e, $l, St $ and $Et$) are defined based on the pattern of time spent in regularly visited places e.g., coffee shops  using their daily activity model in a smart city \cite{do2013places}. $DEC$ and $Rel_i$ are estimated based on the energy usage model of the IoT device.  $I$ and $ Tsr$ are defined based on the specifications of the provided energy services.


\paragraph*{Definition 3} 
\textit{Crowdsourced IoT Energy request}
is defined as a tuple $R=< t_s, t_e, l, RE>$ where: \begin{itemize}[ noitemsep,nosep,leftmargin=10pt,labelsep=1pt,itemindent=0pt, labelwidth=*]
    \item[--] $St$ refers to the timestamp when the energy request is launched. 
    \item[--] $Et$ represents the end time of the period of time, an energy consumer may wait for charging. 
    \item[--] $ l $ refers to the location of the energy service consumer. We assume that a consumer's location is fixed  after launching the request. 
    \item[--] $RE$ represents the required amount of energy. We also assume that the required energy is estimated based on an energy consumption model of the IoT device. 
    
\end{itemize}

\paragraph*{Definition 4} 
\textit{Fairness}
is defined as a function that quantifies the \textit{satisfaction} among a set of requests in a predefined microcell within a predefined time frame. The function takes as an input the available energy $AE$ and the existing requests $R$ and outputs the provisioning fairness score $Fp$ based on the satisfaction $Sf$ of all requests according to their allocated energy $Al$.
\paragraph*{Assumptions}

\begin{itemize}[ noitemsep,nosep,leftmargin=10pt,labelsep=1pt,itemindent=0pt, labelwidth=*]
    
    \item All IoT energy services and requests are deterministic and stationary, i.e., there is an a-priori knowledge about service availability, their QoS values, energy requests, their time constraints, and their demands \cite{Previouswork11}.
    \item The IoT coordinator (at the edge) is responsible for {\em batching} the energy requests from all consumers in a microcell {\em over a predefined period of time}.
    \item The IoT coordinator is also responsible for {\em aggregating} all available energy from all providers in a microcell {\em over a predefined period of time}.  
    \item The energy services may deliver energy to
    \textit{multiple} consumers at the same time without any loss.
\end{itemize}




\begin{figure}[!t]
\centering
\setlength{\abovecaptionskip}{-1pt}
\setlength{\belowcaptionskip}{-18pt}
\centering
\includegraphics[width=.65\linewidth]{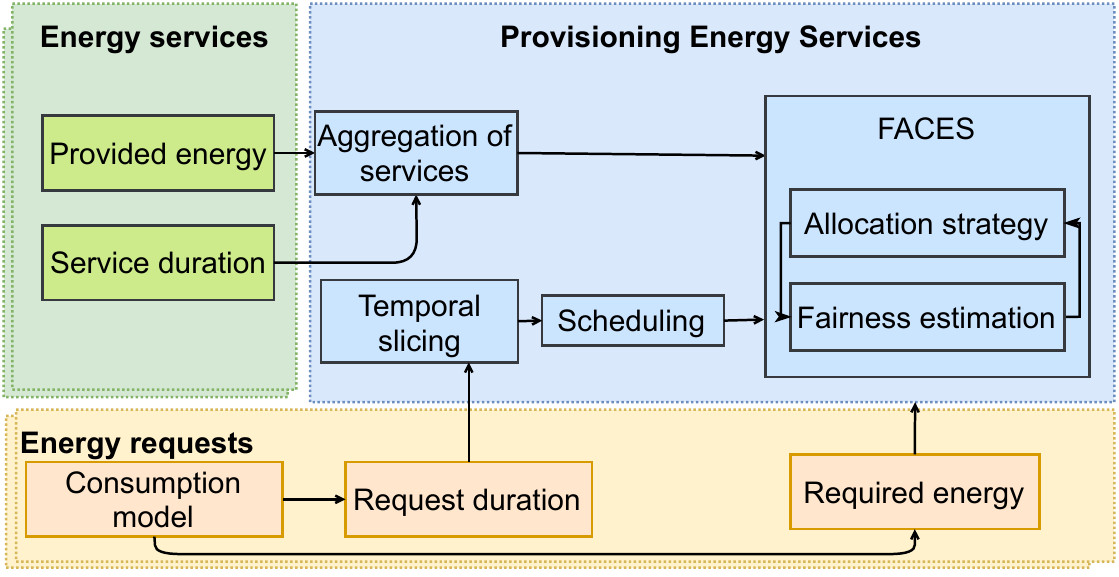}
\caption{ FACES framework}
\label{fig:FRMWRK}
\end{figure}

\vspace{-10pt}
\section{Provisioning energy services}
\vspace{-5pt}
Fig. \ref{fig:FRMWRK} presents the building blocks of our proposed framework. The framework takes as input, the advertised services and the energy requests. Service providers use the previously defined energy service model to advertise their wireless energy services \cite{Previouswork11}. We assume that energy consumers define their energy requirements and their charging waiting time (i.e., request duration) based on predefined consumption models. 

In a microcell $C$, the fairness-aware service provisioning framework is executed at the level of the IoT coordinator in the edge (i.e., a router within the microcell $C$).  Given a set of crowdsourced IoT energy services in the microcell $C$, $ S=\{ S_{1}, S_{2}, \dots S_{n} \} $ and a set of all existing requests within the same time interval $W$, $R = \{ R_{1}, R_{2}$, $\dots R_{m} \}$. The IoT coordinator aims at minimizing the wastage $Wsg$ while provisioning the aggregated energy to all existing requests within the time window $W$ by performing the following steps:
\begin{figure}
\centering
\setlength{\abovecaptionskip}{-2pt}
\setlength{\belowcaptionskip}{-20pt}
 \includegraphics[width=0.85\textwidth]{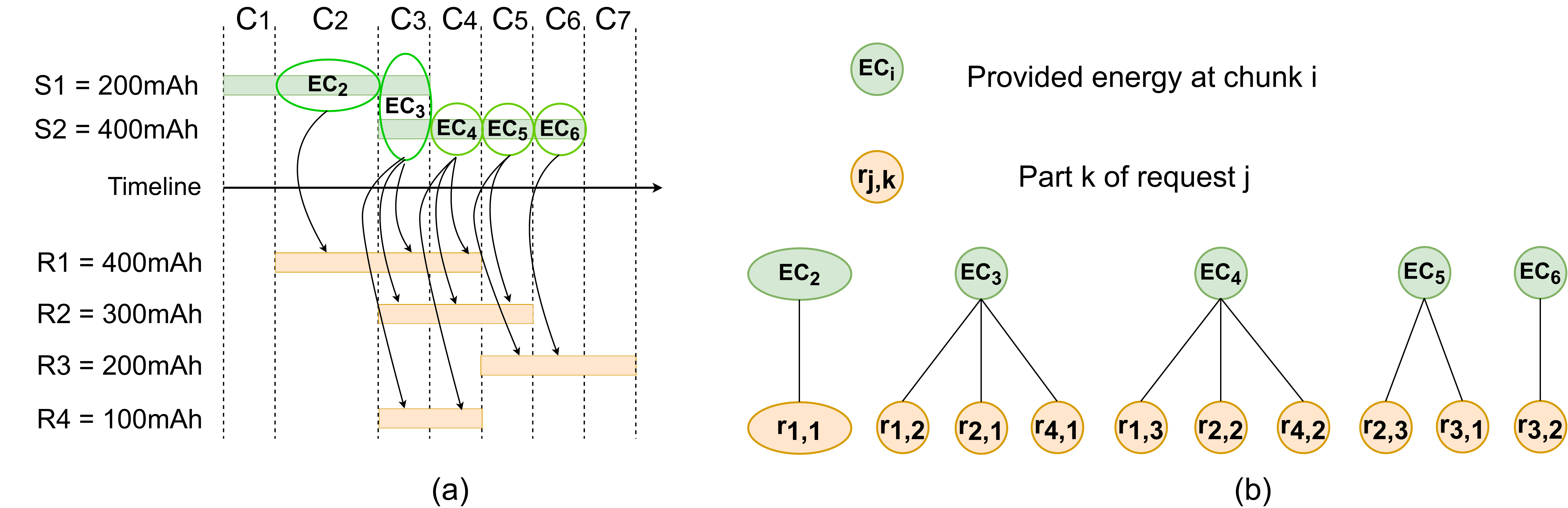}
\caption{\small (a) Chunking crowdsourced IoT energy services and requests (b) The bipartite graph representation of the time-constrained allocation problem}
    \label{fig:prov_Frmwrk}
\end{figure}
\subsubsection{Temporal slicing}
The goal of the temporal slicing module is \textit{to segment} the requests time intervals and define the overlapping parts. We follow the temporal chunking of energy services in \cite{Previouswork11} to define the time slices of the time window $W$. We define all the possible timestamps where a preemptive scheduler may switch to another request. Each timestamp is either the start time or the end time of existing requests. We divide the time window $W$ into several time slices based on these timestamps. The time slots represent the arrival time of a new request or the exit time of an existing request (i.e., vertical dotted lines in Fig. \ref{fig:prov_Frmwrk} (a)) \cite{Previouswork11}. For example, $S_2$  in Fig. \ref{fig:prov_Frmwrk} may be {\em temporally} chunked into four {\em parts of services} $S_iP_j$ (i.e., part $i$ of service $j$) as follows: $<S_2P_1$, $S_2P_2$, $S_2P_3$, $S_2P_4>$. 
Similarly, $R_1$ in Fig. \ref{fig:prov_Frmwrk} may be {\em temporally} chunked into three {\em parts of requests} $R_iP_j$ as follows: $<R_1P_1$, $R_1P_2$, $R_1P_3>$. 
At each chunk, a part of service $S_iP_k$ may provision a part of a request $R_iP_j$ within the same chunk. For example, in Fig. \ref{fig:prov_Frmwrk} (b), $S_1P_2$ provides energy to $R_1P_1$ since they are within the same chunk. Each request $R_i$  may be provisioned by a {\em composition} of parts of services $S_iP_k$ where a part of service delivers energy to a part of the request at each chunk \cite{lakhdari2020composing}. After scheduling the requests, it is more challenging to provide energy to multiple requests in a fair way.


\subsubsection{Aggregation of services}
The framework starts by \textit{aggregating} all the available services within the time window $W$. Energy services are composed according to their spatio-temporal features. We use the  framework of composing crowdsourced IoT energy services proposed by Lakhdari et al. \cite{lakhdari2020composing}. The composition considers the time interval of each service to define a composite energy service that includes all the available services. The IoT coordinator defines a sequence of time intervals and the available energy at each time interval. If two or more energy services overlap within a time interval, the IoT coordinator sums the provided energy by all services available at that time interval.

$$AggE = \sum_i DEC_i ~ \forall CES_i\in agt$$.
$agt$ represents the composite energy service resulting of the spatio-temporal composition of available services $S$. $AggE$ is a QoS of the composite energy service $agt$. $AggE$ denotes the total energy provided by the aggregated services.
\subsubsection{Scheduling} 
The scheduling module takes the segmented requests and starts by planing the provision for only the non overlapping segments for each request. The allocated energy amount $Al_k$ to a request $R_k$ can only be provided from the available energy within its time interval [$St_k, Et_k$]. $[S_k, E_k]$  represents the time interval when a request $R_k$ may receive energy. $Av_k$ represents the available energy within the time interval $[S_k, E_k]$. For example, in figure \ref{fig:prov_Frmwrk} the IoT coordinator provides energy to $R_4$ only from the available energy within the time chunks  $C_3$ and  $C_4$ . 
\subsubsection{Fairness estimation} 
In energy provisioning, the fairness score $Fp$ is calculated based on the sparsity in allocating energy to all existing requests. Intuitively, less allocation sparsity among requests reflects more fairness. We define a sparsity function $Sf$ to estimate the sparsity based on the allocated energy $Al_k$ to each request $R_k$ from the available energy $Av_k$ within its time interval. The fairness estimation module initially calculates the fairness score after only provisioning the non-overlapping requests segments. The heuristic cannot estimate the fairness and allocate energy to the overlapping request segments only if there is a prior knowledge about the allocated energy to the non-overlapping segments. 
\subsubsection{Allocation algorithm}
The allocation algorithm aims at minimizing the wastage of the aggregated energy and maximize the fairness among requests. We transform the problem of fairness-aware energy provisioning into a time-constrained resource allocation problem as follows:
\begin{equation*}
\begin{aligned}
& {\text Minimize} & & wsg = AggE-\sum_{i=1}^{n} Al_i\\
& {\text Maximize} & & Fp = sf(Al_i), \forall R_i\in R\\
& {\text Subject~to} & & [St_i, Et_i]=[S_i, E_i], \forall R_i\in R\\
& {\text Where}  & & \forall R_i\in R\\
&  & & {\text [St_i, Et_i]~ is~ the ~ interval ~of~ the ~request ~R_i}\\
&  & & {\text [S_i, E_i]~ is ~the ~interval~ when ~ R_i ~ may ~receive ~energy}\\
\end{aligned}    
\end{equation*}
The algorithm aims at solving the said \textit{multi-objective optimization} by efficiently provisioning the overlapping segments of requests. Next, the \textit{fairness estimation module} recalculates the fairness score at each optimization step. In the following, we explain the fairness concept for energy requests in a crowdsourced IoT energy market as we present the building blocks of the heuristic-based allocation algorithm.

\section{Fairness-Aware Crowdsourcing of Energy Services (FACES)}\label{chunkingggg}
In a framework for provisioning energy services, there is a dual need to, on the one hand, maximize energy use from a consumer perspective and, on the other hand, maximize the provisioning of energy from the providers' point of view. We propose a scheme whereby energy requests from all consumers in a microcell (over a predefined time frame) are batched while all available energy from all providers in a microcell is aggregated. Fig. \ref{fig:MTVScen} presents a batched set of energy services and requests from 5:00 to 6:30. A global view of all available services and requests within a predefined time interval allows the IoT coordinator to aggregate the provided energy by \textit{composing} all services based on their availability time intervals \cite{lakhdari2020composing}. The coordinator aims to efficiently and fairly share the aggregated energy among all the existing requests according to their time constraints.

\subsection{Fairness estimation}
Fairness-aware provisioning does not necessarily imply an equal allocation of the available energy to all requests.  In a crowdsourced IoT energy market, distributive fairness \cite{leventhal1980should} is defined by equally provisioning requests according to their {\em features} (i.e., requirements and time constraints). Provisioning IoT energy services fairly aims to {\em satisfy} more consumers rather than maximizing the energy allocation for some requests. Typically in resource allocation problems, when the resources are limited, consumers will not be satisfied by an efficient allocation of available resources to all requests \cite{radunovic2007unified}. However, in a crowdsourced IoT environment, fairness-aware provisioning of energy services is claimed to increase the utilization of the available energy services. We rely on the satisfaction of energy consumers to monitor the fairness of our service provisioning framework.
\subsubsection{Satisfaction:} 

We first define a satisfaction score $Sf$ for energy consumers. Intuitively, the amount of the acquired energy is directly proportional to the satisfaction of consumers. However, consumers already realize the limited availability of energy services in the crowdsourced IoT energy market. In this work, we consider an {\em altruistic} behavior of energy consumers. Consumers' goal is both to maximize their allocated energy and to contribute selflessly to fair provisioning. They may adjust their satisfaction score based on the market (i.e., available energy services and existing requests). 

\paragraph*{Definition 5} \textit{The satisfaction $Sf_i$} of an energy consumer toward a request $R_i$ reflects their perception of the allocated energy $Al_i$  to their request. We quantify $Sf_i$ score for a request $R_i$ based on the allocated energy $Al_i$ and the available energy in the crowdsourced market as follows:

\[
    Sf_i= 
\begin{cases}
    \frac{Al_i}{RE_i}\times \frac{\sum_{i=1}^{n}Al_i}{\sum_{j=1}^{m}DEC_j},& \text{if } Al_i\leq RE_i\\
    1,              & \text{otherwise}
\end{cases}
\]

Where $\sum_{i=1}^{n}RE_i$ represents all the requested energy in the crowdsourced market by the set of all existing requests $R_i \in ExR$. $\sum_{j=1}^{m}DEC_j$ represents all the energy in the market provided by the available energy services $S_j \in AvS$
\subsubsection{Fairness score:}
We define a global fairness metric for energy services provisioning $Fp$ based on the satisfaction of all consumers. We measure the global fairness of the service provisioning framework by estimating the overall proximity score among the satisfaction scores of all consumers \cite{basik2018fair}. An unfair provisioning plan is reflected by {\em sparse} satisfaction scores among consumers (i.e., some requests have high satisfaction score than others). Contrarily, less sparse satisfaction scores reflect higher proximity among all requests.

\begin{figure}[!t]
\setlength{\abovecaptionskip}{-1pt}
\setlength{\belowcaptionskip}{-18pt}
\centering
\includegraphics[width=.55\linewidth]{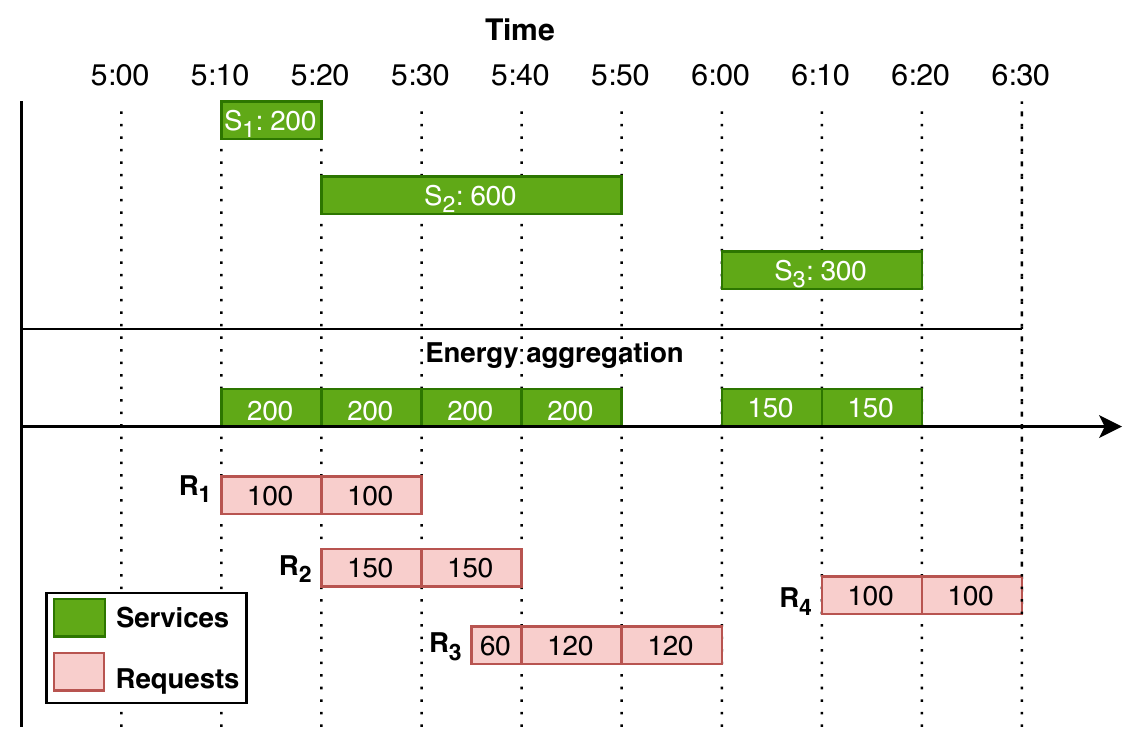}
\caption{ Example of batched energy services and requests}
\label{fig:MTVScen}
\end{figure}

\paragraph*{Definition 6} \textit{The global fairness} $Fp$ is a metric to quantify the sparsity of the satisfaction scores  among all requests in the crowdsourced IoT energy market. We capture the global fairness $Fp$ using the {\em information entropy} \cite{shannon1948mathematical}. The information entropy  measures the disorder degree of all requests $R_i \in ExR$  based on their satisfaction score$Sf_i$ as follows:   
$$Fp(ExR) = -\sum Sf_i\log_2 Sf_i$$

\begin{algorithm}[t!]
\footnotesize
    \renewcommand{\algorithmicrequire}{\textbf{Input:}}
    \renewcommand{\algorithmicensure}{\textbf{Output:}}
    \caption{Heuristic-based fairness-aware energy allocation}
    
    \label{alg:STcompo}
    \begin{algorithmic}[1]
        \Require
         $C$, $ S=\{ S_{1}, S_{2}, \dots S_{n} \} $, $W<St,Et>$, $R = \{ R_{1}, R_{2}$, $\dots R_{m} \}$.
        \Ensure $Al = \{ Al_{1},Al_{2}$, $\dots Al_{m} \}$.
        \Statex \text{ // Aggregating energy services}
        \For{$ S_i \in S $}
            \State $AggE = \sum_i DEC_i ~ \forall CES_i\in agt$
        \EndFor
        \Statex \text{ // Chunking energy requests}
		\State $Chunk_{0}.st \gets W.St $
		\For{$ int~t= W.St~to~ W.Et $}
		    \If {$(\forall~R_i\in R~and~t =R_{i}.st ~or~t = R_{i}.et)$}
		        \State $ Chunk_{i}.et\gets t $
		    \EndIf
		    \Statex \text{ // create new chunk}
		    \If {$t \neq W.Et  $}  
		        \State $ Chunk_{i+1}.st\gets t $
		        \State $ t \gets t+1 $
		    \EndIf
		\EndFor
        \For{   $R_i\in R$}
        \Statex \text{// Chunk-based provisioning}
        \For{   $Ch\in Chunk$}
        	\Statex \text{// First, provision non-overlapping requests}
		    \Statex \text{// Second, provision overlapping requests per chunk}
		    \Statex \text{// $Ch_pr$ is the set of  partial requests within a chunk}
		    \Statex \text{// $Ch_AE$ is the available energy within a chunk}
		    \For{   $Pr\in Ch_pr$}
		        \State $pr \gets Ch_AE/|Ch_pr|$
		    \EndFor
		    \State \textbf{While}\text{($pr_i \in R_i$)}
		        \State    \text{$Al_i \gets Al_i \cup \{ pr_i\}$} 
 		    \State \textbf{End While}
        \EndFor
        \State \text{$Al \gets Al \cup \{ Al_i\}$}
        \EndFor
        \State \Return $ Al $
    \end{algorithmic}
\end{algorithm}

\subsection{Heuristic-based fairness-aware allocation algorithm}
We propose a heuristic-based allocation strategy, i.e., Fairness-Aware Crowdsourcing of Energy Services (FACES), which extends the traditional resource allocation strategies by optimizing the energy allocation for \textit{overlapping} requests. Our proposed heuristic does not only consider the allocated time for each request. It also considers sharing the available energy when two or multiple requests are overlapping. For example, at the time segment [5:20, 5:30] in Fig. \ref{fig:MTVScen} when $R_1$ and $R_2$ overlap, FACES divides the available energy at that time segment (200 mAh) between $R_1$ and $R_2$. Algorithm \ref{alg:STcompo}  presents the pseudocode of the heuristic-based fairness-aware provisioning. First, the energy services are aggregated by the IoT coordinator (Line 2). Next, the time interval is chunked based on the arrival time of energy requests (Lines 3-9). the provisioning framework starts by the non overlapping requests. Then, for each chunk containing overlapping requests, the available energy at that chunk would be equally split among those chunks (Lines 10-13). Finally, all the allocated energy is aggregated per request (Lines 14-17).    Provisioning overlapping requests simultaneously improves fairness among requests (i.e., $\sigma=8.03$) and minimizes the wastage of the available energy, i.e., $10\%$ of the available energy. An optimal fairness-aware provisioning plan will maximize the consumption of the available aggregated energy. 

The complexity of the proposed fairness algorithm can be estimated based on the number of available requests and the number of chunks C and the number of overlapping requests at each chunk. The runtime complexity of FACES is O(Cn). If we consider n as the number of available partial requests within a chunk.




\subsection{Assessment of allocation strategies}

\begin{table}[!t]
\caption{ Allocation strategy effect on the energy provisioning}
\label{tab:FairnessPolicies}
\centering
    \setlength{\abovecaptionskip}{-10pt}
    \setlength{\belowcaptionskip}{-25pt}
\small
\resizebox{.6\linewidth}{!}{%
\begin{tabular}{cl|c|c|c|c|}
\cline{3-6}
\multicolumn{2}{l}{}                     & \multicolumn{4}{|c|}{Algorithms' results (\%)} \\ \hline
\multicolumn{2}{|c|}{Request (Capacity)} & FCFS      & RR        & P-FCFS     & FACES    \\ \hline
\multicolumn{2}{|c|}{R1 (200 mAh)}       & 100       & 50        & 100        & 100      \\ \hline
\multicolumn{2}{|c|}{R2 (300 mAh)}       & 0         & 100       & 50         & 83       \\ \hline
\multicolumn{2}{|c|}{R3 (180 mAh)}       & 100       & 67        & 67         & 85       \\ \hline
\multicolumn{2}{|c|}{R4 (100 mAh)}       & 100       & 100       & 100        & 100      \\ \hline \hline
\multicolumn{2}{|c|}{Provision wastage}  & 48.72     & 26.93     & 20.52      & 10       \\ \hline
\multicolumn{2}{|c|}{Provision unfairness} & 43.30     & 21.60     & 21.60      & 8.03     \\ \hline
\end{tabular}%
}
\end{table}

Table. \ref{tab:FairnessPolicies} presents different allocation plans for the aggregated energy to accommodate $R_1$, $R_2$, $R_3$, and $R_4$ (see Fig. \ref{fig:MTVScen}). We calculate the amount of energy (i.e., capacity) each request can receive energy according to different allocation strategies. 
The first allocation plan follows the FCFS (First Come First Served) scheduling strategy. Each request is provided with energy according to its arrival (i.e., start time). Contrarily, the Round Robin (RR) is a preemptive scheduler that allocates a fixed time interval for each request. In our example, we consider 10 minutes as a fixed time interval. RR reduces the provision wastage compared to FCFS. A Preemptive implementation of FCFS (P-FCFS) extends the FCFS strategy by considering overlapping requests. FCFS, RR, and P-FCFS strategies are runtime efficient schedulers that consider only one request at a time. Their goal is to allocate time to each request \textit{efficiently}.
We evaluate the outcome of the different energy allocation strategies based on the energy \textit{wastage} and \textit{fairness} among requests. We define energy wastage as the amount of lost energy that could not be utilized to fulfill the capacity of all requests. For example, the Round Robin algorithm in table \ref{tab:FairnessPolicies} exhibits 26.93\% wastage. For a simplistic illustration of fairness, we use the standard deviation $\sigma$ to estimate the fairness. Intuitively, a better fairness is reflected by a lower value of $\sigma$.  In the example illustrated by Fig. \ref{fig:MTVScen}, P-FCFS provides more fairness compared to FCFS because the provision sparsity of P-FCFS ($\sigma=21.60$) is less than the one of FCFS.

\section{Experiments}
We conduct a set of preliminary experiments to evaluate the proposed theoretical concepts of fairness-aware provisioning of crowdsourced energy services. We essentially assess the \textit{effectiveness} of different fairness strategies on maximizing the utilization of the available energy within a microcell. We measure the ratio of the consumed energy over the available energy across different microcells. We monitor the changes in the fairness score and the energy utilization ratio while varying the number of energy requests. We implement two variants of the proposed approach(FACES) and compare them  with traditional resource allocation algorithms, namely, FCFS, P-FCFS, and Max-min fair scheduling \cite{tassiulas2002maxmin}.

\subsection{Dataset and experiment environment}

We create a crowdsourced IoT environment scenario close to reality. We mimic the energy sharing behavior of the crowd within microcells by utilizing a dataset published by IBM for a coffee shop chain with three branches in New York city\footnote{https://ibm.co/2O7IvxJ}. The dataset consists of transaction records of customers purchases in each coffee shop for one month. Each coffee shop consists of, on average, 560 transnational records per day and 16,500 transaction record in total. We use the IBM dataset to simulate the spatio-temporal features of energy services and requests. The dataset contains information about the crowd's behavior in coffee shops. People may check-in, rate, and recommend these venues. In our experiment, we only focus on people's check-ins information. We extract the crowd size for each coffee shop at each hour ($hour$) of the day ($weekday$). We assume these people as IoT users. They may offer energy services from their wearables while staying in the coffee shop. We define spatio-temporal features of energy services by generating customers' check-in and check-out timestamps to confined areas using the previously extracted data from their transactions. For example, the start time $st$ of an energy service from an IoT user is the time of their check-ins into a coffee shop. Energy request time $R.st$ and duration $R.et$ are also generated from check-in and check-out times of customers. \textit{To the best of our knowledge, it is challenging to find a dataset about the wireless energy transfer among human-centric IoT devices}. We use a random uniform distribution to generate the energy amount for each request and the amount of provided energy for each service.

\begin{figure}[!t]
\centering
\begin{minipage}{.5\linewidth}
  \centering
  \includegraphics[width=\linewidth]{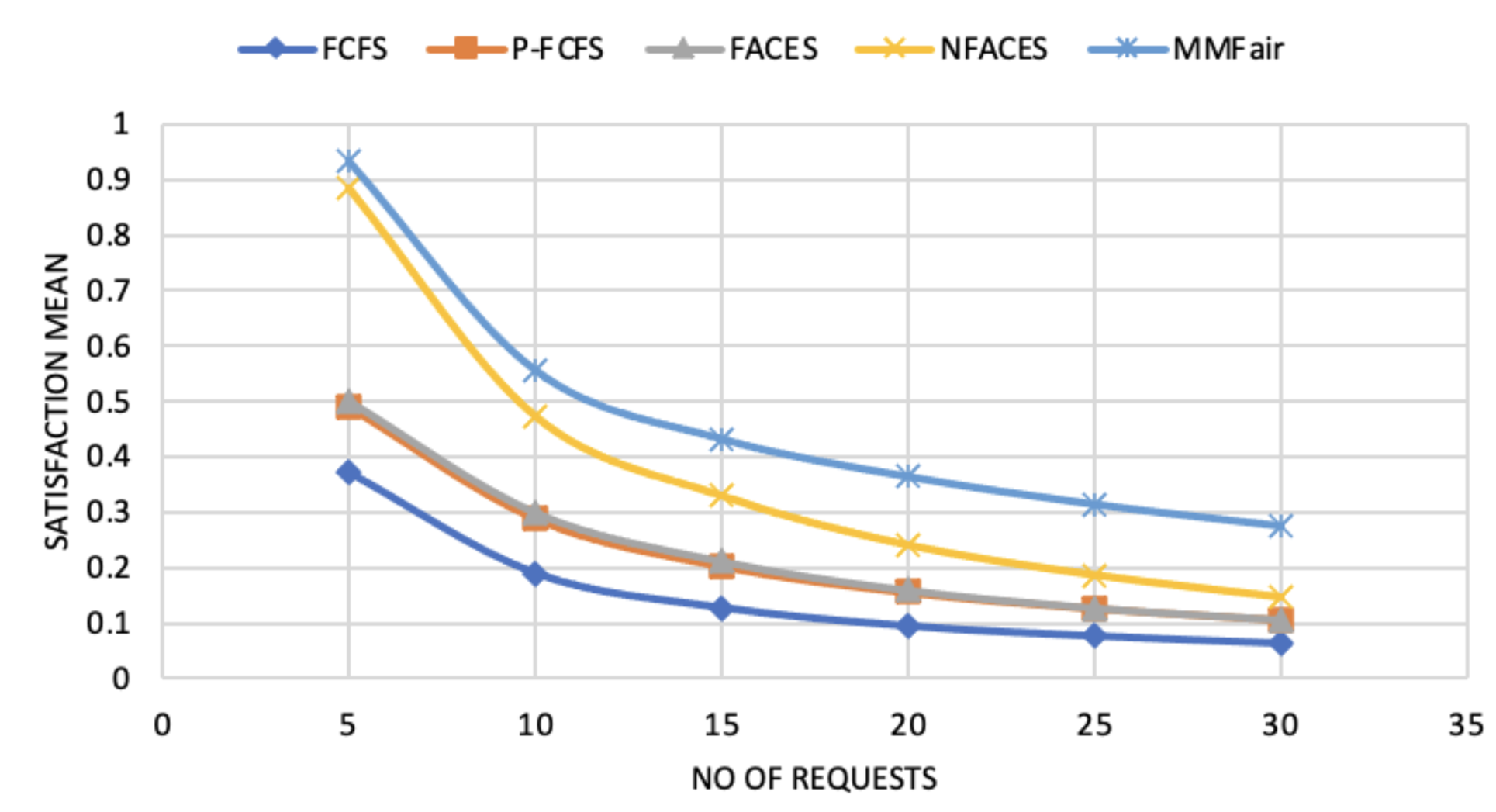}
   \setlength{\abovecaptionskip}{-10pt}
   \setlength{\belowcaptionskip}{-15pt}
  \caption{The mean score of consumers satisfaction. }
\label{rs1}
\end{minipage}%
\begin{minipage}{.5\linewidth}
  \centering
  \includegraphics[width=\linewidth]{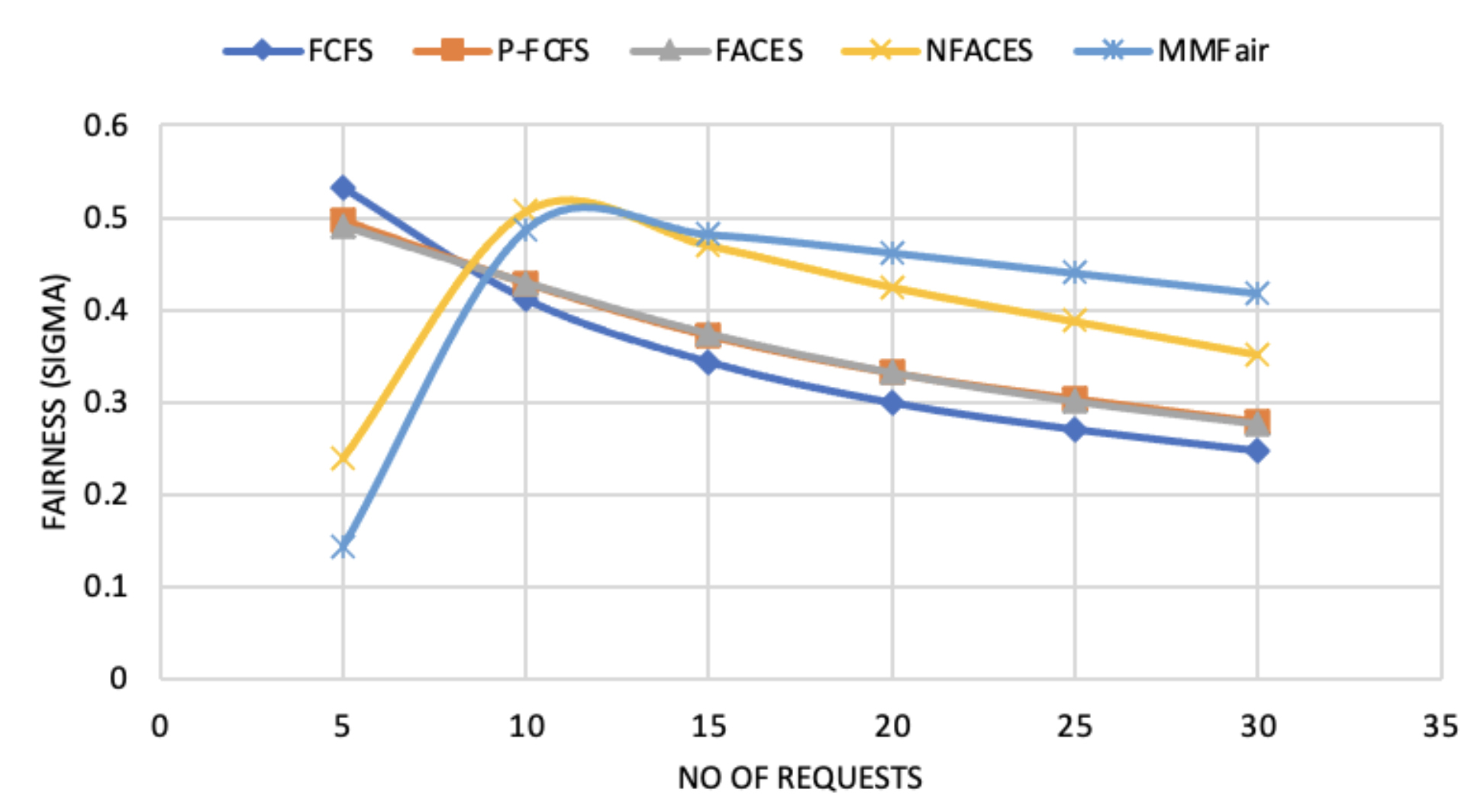}
  \setlength{\abovecaptionskip}{-10pt}
  \setlength{\belowcaptionskip}{-15pt}
  \caption{The standard deviation score of consumers satisfaction. }
\label{rs2}
\end{minipage}
\end{figure}

\begin{figure}[!t]
\centering
\begin{minipage}{.5\linewidth}
  \centering
  \includegraphics[width=\linewidth]{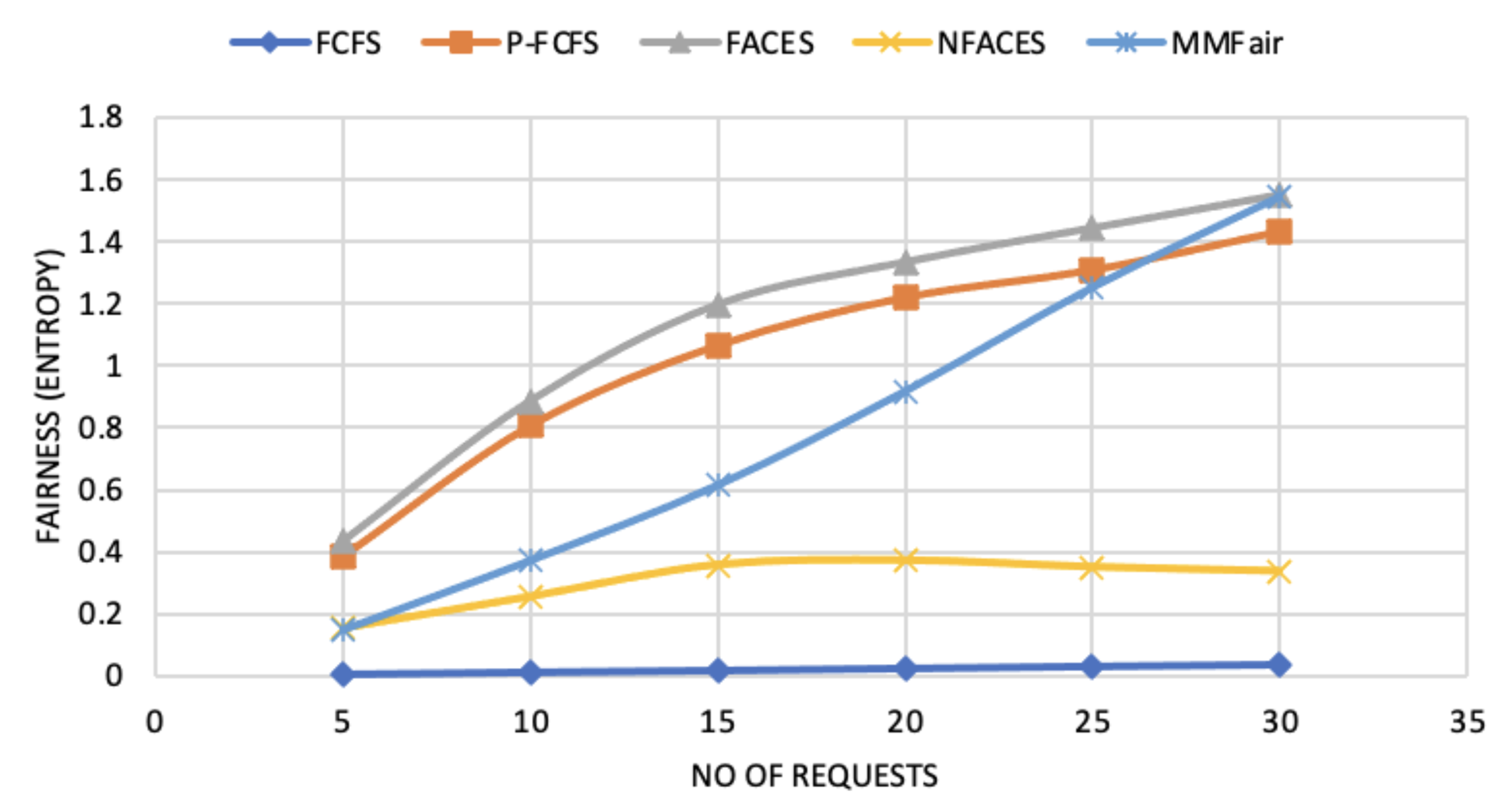}
   \setlength{\abovecaptionskip}{-10pt}
   \setlength{\belowcaptionskip}{-15pt}
  \caption{The Entropy score of consumers satisfaction. }
\label{rs3}
\end{minipage}%
\begin{minipage}{.5\linewidth}
  \centering
  \includegraphics[width=\linewidth]{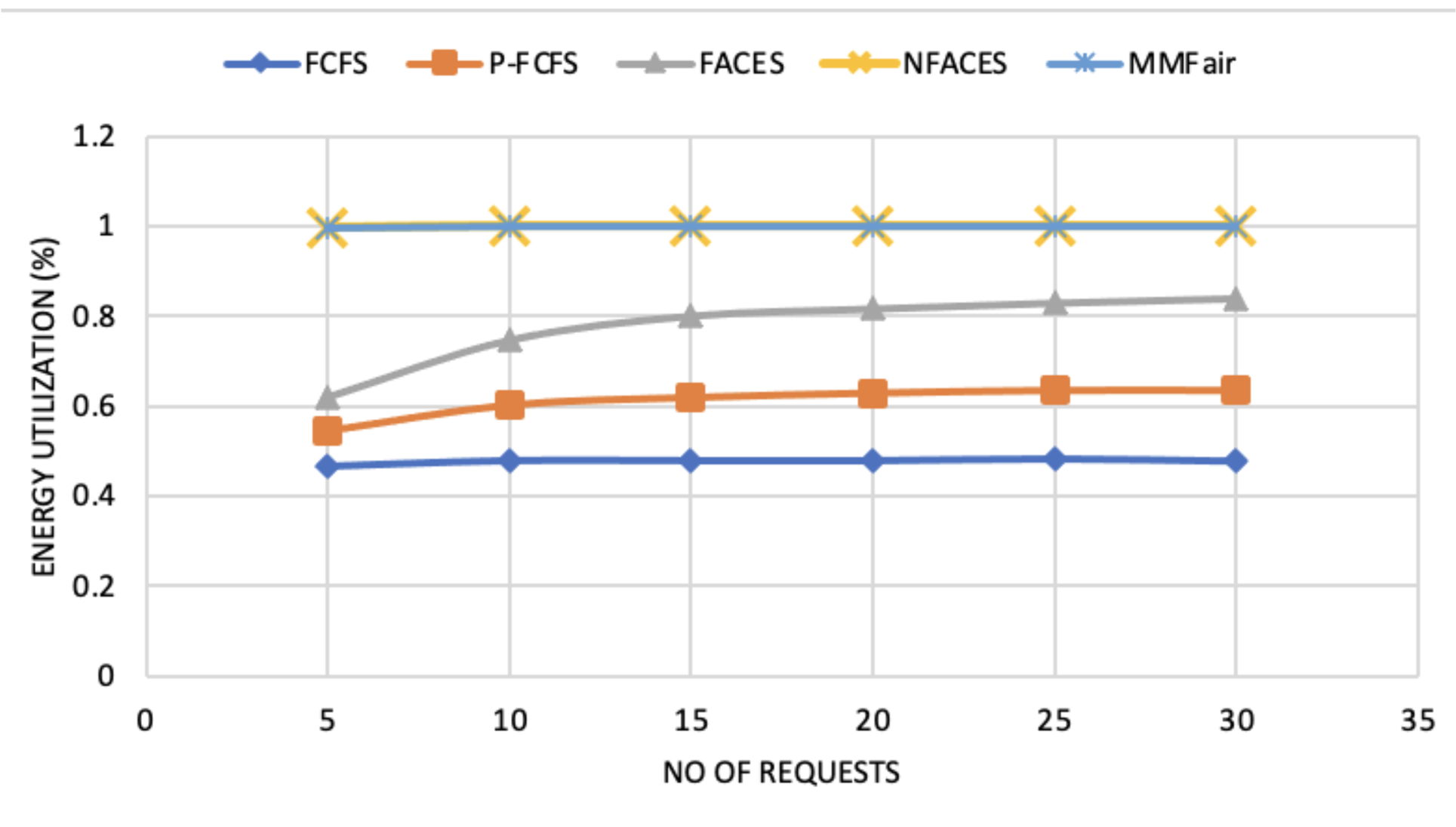}
  \setlength{\abovecaptionskip}{-10pt}
  \setlength{\belowcaptionskip}{-15pt}
  \caption{The energy utilization ratio }
\label{rs4}
\end{minipage}
\end{figure}

\subsection{Effectiveness}
We implement two variants of the FACES framework, namely, FACES and NFACES. FACES considers only one request at a time, similar to FCFS and P-FCFS. NFACES, however, considers multiple requests at a time similar to max-min fair scheduler \cite{tassiulas2002maxmin}. FACES and NFACES chunk the requests before provisioning. The chunks are defined based on the overlapping between requests \cite{lakhdari2020composing}. We implemented a modified version of the Max-min scheduler to consider the temporal constraints of energy requests. For each chunk, if there is more than one request, max-min fair scheduling is performed. Contrarily, to the max-min fair scheduler, NFACES privileges the partial requests with the highest required amount at each chunk. The remaining available energy at that chunk is reallocated to the remaining requests in descending order. In what follows, we assess the fairness metrics, sigma and entropy for all the algorithms along with their performance in terms of the energy utilization. 

\subsubsection{Fairness}
We evaluate the effectiveness of the proposed framework by assessing the effect of the fairness-aware allocation strategy on energy utilization. We first investigate fairness through different metrics, namely, the mean, standard deviation, and entropy satisfaction score for existing requests. Fig. \ref{rs1} illustrates the change of the satisfaction mean value which reflects the average of acquired energy amount per request. Intuitively, the more requests, the less energy amount to acquire per request for all the five allocation strategies. The following figure, Fig. \ref{rs2} presents the dispersion of the energy requests satisfaction score around the mean. This metric reflects the variation of the satisfaction score among requests. With a larger number of requests (more than 10 requests), the acquired energy decreases significantly, which explains the decrease and the convergence of the standard deviation due to the decrease of the satisfaction score among most of the requests.

The information entropy captures the multi-modal dispersion and irregularities in the distribution of the satisfaction score of energy consumers (see Fig. \ref{rs3}). We leverage the information entropy to monitor the fairness in provisioning energy requests. A lower value of entropy means better fairness in the allocated energy. It is worth mentioning that the entropy metric could capture the small variations in the satisfaction score when the number of requests is larger. These variations cannot be noticed only with a fairness metric based on the standard deviation (see Fig. \ref{rs2}). In Fig. \ref{rs3} FCFS exhibits a near zero score for the entropy, which can be explained by the fact that most of the requests satisfaction score is equal to zero. NFACES, however, demonstrates a better performance behavior in terms of fair provisioning for energy requests. With a larger number of requests, the entropy values for NFACES are lower than those of P-FCFS and Max-min fair scheduler. 
\vspace{-10pt}
\subsubsection{Energy utilization}
The fourth experiment compares the Energy Utilization (See Fig. \ref{rs4}). The goal of proposing a fairness-aware provisioning framework is to leverage fairness as a driver to increase the utilization of the available energy services in a crowdsourced IoT environment by increasing the participation of energy consumers. Energy utilization is the ratio of the amount of the allocated energy services over the total amount of available energy services. Overall, all the energy allocation techniques converge after 15 requests. FACES exhibits the best performance behavior among the three algorithms that consider only one request at a time (i.e., FCFS,P-FCFS, and FACES). The energy utilization ratio is significantly higher with NFACES and Max-min fair scheduler, an expected behavior by these two strategies as a result of considering overlapping requests, i.e., more than one request at a time. 

In conclusion, this set of preliminary experiments confirms our claim that fairness-aware allocation strategies would better utilize the available energy in a crowdsourced IoT environment. It is worth mentioning that NFACES exhibits far better fairness behavior compared to Max-min fair scheduler, nonetheless the same performance in terms of the energy utilization.


\section{Related work}

Service computing is a key enabler for wireless energy sharing. {\em Service composition} is expected to play a vital role in the crowdsourced IoT environment. A single IoT energy service may not fulfill the requirement of a consumer due to the limited resources of IoT devices \cite{lakhdari2020Vision}. Several service composition techniques have been proposed. Mainly, the service composition techniques can be categorized into a functionality-based composition or QoS-based composition. For example, Tan et al.  \cite{tan2010data} proposed a data-driven composition approach that uses Petri-nets to meet the application’s functional requirements.  Wang et al.  \cite{wang2014constraint} address the problem of service functionalities constraints by introducing a pre-processing technique and a graph search-based algorithm to compose services.

Service selection and composition also play an important role in emerging fields such as cloud computing, IoT-based smart systems \cite{chaki2020fine}\cite{shahzaad2021resilient}. In IoT, services are mainly composed according to their spatio-temporal features \cite{hamdi2021spatiotemporal}. They also must fulfill consumer preferences (QoS). For example, Lakhdari et al. design and implement a spatio-temporal service composition framework for crowdsourced IoT services \cite{Lakhdari2021Proactive}. User preferences are used to define the spatial and temporal composability models, Neiat et al. proposed a spatio-temporal service composition framework to describe and compose region services like WiFi hotspots \cite{neiat2017crowdsourced}. Existing energy service composition frameworks mainly consist of the real-time discovery and selection of nearby energy services \cite{lakhdari2020fluid}. The focus of these composition techniques was only on the spatio-temporal composability \cite{lakhdari2020composing} and addressing the challenges related to the energy fluctuation and the mobility of the available services \cite{lakhdari2020Elastic}.

The current work adds a new contribution to the field of energy crowd sharing. Indeed,  existing service composition techniques address the challenges related to one single consumer at a time. Our proposed approach considers provisioning multiple consumers in a predefined time and space. To the best of our knowledge, the work is among the first attempts to address fairness challenges in a crowdsourcing IoT energy services.
\vspace{-5pt}
\section{Conclusion}
\vspace{-5pt}

We proposed a fairness-aware framework for provisioning IoT energy services in a crowdsourced IoT environment. We introduced the concept of {\em fairness} to efficiently provision available IoT energy services and accommodate multiple energy requests in a microcell within a predefined time frame. The under-provision of energy requests may demotivate consumers to participate in the crowdsourced IoT energy market. We investigated different allocation strategies to provision energy services, namely, FCFS, P-FCFS, and Round Robin. We defined a fairness model based on the satisfaction of consumers. Our goal is to leverage the fairness as a means to maximize the utilization of the available energy services. We designed and develop a fairness-aware scheduling framework to provision IoT energy services. We conducted a set of preliminary experiments to assess the effectiveness of the proposed framework.
\section*{Acknowledgment}
This research was partly made possible by DP160103595 and LE180100158 grants from the Australian Research Council. The statements made herein are solely the responsibility of the authors.
%
%
\bibliographystyle{splncs04}
\bibliography{mainFile}

\end{document}